\documentclass[aps,prb,twocolumn,groupedaddress,showpacs]{revtex4}
\usepackage{graphicx}
\newcommand{\Vec}[1]{\mbox{\boldmath$#1$}}

\begin{document}


\title{Tight-binding photonic bands in metallophotonic waveguide networks
and flat bands in kagome lattices}


\author{Shimpei Endo}
\email{endo@cat.phys.s.u-tokyo.ac.jp}
\author{Takashi Oka}
\author{Hideo Aoki}
\affiliation{Department of Physics, University of Tokyo, Hongo, Tokyo 113-0033, Japan}


\date{\today}
\begin{abstract}

We propose that we can realize ``tight-binding photonic bands" 
in metallophotonic waveguide networks, where the photonic bound states localized around the crossings of a network form a tight-binding band.  
The formation of bound states at the crossings is distinct from 
the conventional bound states at defects or virtual bound states in photonic crystals, but 
comes from a photonic counterpart of the zero-point 
states in wave mechanics. 
Model calculations show that the 
low-lying photon dispersions are indeed described accurately by the tight-binding model. 
To exemplify how we can exploit the 
tight-binding analogy for {\it designing} of photonic bands, we propose a ``flat photonic band" in the kagome network, in which the 
photonic flat band is shown to arise with 
group velocities that can be as small as 1/1000 times the velocity of light in vacuum. 
\end{abstract}

\maketitle

Photonic crystals (PhCs) are interesting because they enable 
us to control optical properties 
just as we can manipulate properties of electrons in solids.\cite{pendry1996cpb} PhCs have attracted both experimental\cite{PhysRevLett.67.3380,wanke1997lrp} and theoretical\cite{PhysRevLett.58.2486,PhysRevLett.69.2772,PhysRevLett.94.073903} interests, and the dramatic developments in the field in fact owe much to the analogies between PhCs and solids.  
Many phenomena first found in solids have also been found in PhCs, such as the appearance of 
band gaps\cite{PhysRevLett.58.2059,PhysRevLett.67.2295}, Anderson localization,\cite{PhysRevLett.58.2486,PhysRev.109.1492} modes localized to defects in the band gap frequency,\cite{PhysRevLett.67.3380} and lights with small group velocities or heavy photons.\cite{PhysRevLett.94.073903}  

\begin{figure}[b]
 \includegraphics[width=5.5cm]{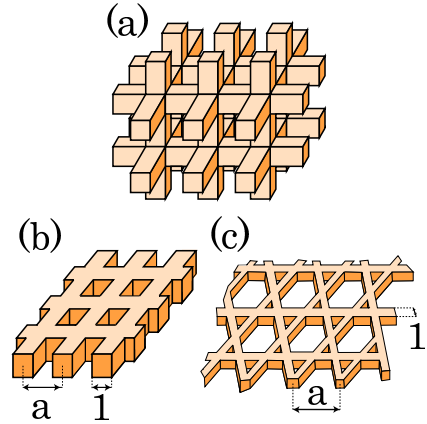}
 \caption{\label{MPWN}(Color online) 
 Metallophotonic waveguide networks are schematically shown: (a) a three-dimensional cubic lattice, (b) a two-dimensional square lattice, and (c) a two-dimensional kagome lattice.  In each case, light is confined in the shaded regions, while the rest of the space is filled with a light-prohibiting material such as a metal. For two-dimensional cases as in (b) and (c), the system is assumed to have a large extension in the $z$ direction.
}
 \end{figure}

While photonic bands realized in periodically modulated dielectric media have 
been extensively studied, there should be further avenues that have not been thoroughly explored.  In the 
electronic band structures, 
there are two opposite approaches well known in solid-state physics: the nearly-free-electron model and the tight-binding model (TBM).  
While the former is perturbative, the TBM shows a great variety of band structures, and numerous lattices have been analyzed in solid state physics.  
In this Brief Report we pose a question: can we realize tight-binding photonic bands?  
Here, we propose a way to realize that.  We predict that 
metallophotonic waveguide networks (MPWNs), where waveguides are joined together to form a network structure as schematically depicted in Fig. \ref{MPWN}, 
should realize the tight-binding photonic bands through an unsuspected 
``zero-point" localized photonic states. The network is taken to be surrounded by a material into which light cannot penetrate, so that the light is confined inside the network. An obvious material is a metal for a sufficiently smaller frequency than the plasma frequency, such as in the microwave regime. Networks of line-defect waveguides in PhCs may be another possibility, 
but the detail of the surrounding material is unimportant as long as the confining property and low-loss condition are satisfied.  



We note that there exist literatures that propose some ways to realize tight-binding photonic bands for coupled waveguides or cavities in PhCs,\cite{amiri2006tba,PhysRevB.57.12127,PhysRevLett.84.2140} and calculation methods based on the TBM have been successfully applied to these systems.\cite{PhysRevLett.81.1405,albert2002pcm} However, most tight-binding studies have relied on the introduction of defects, 
around which a localization mechanism inside the band gap is sought. 
The MPWN we propose here, by contrast, exploits a different mechanism. The tight-binding photonic bands in MPWNs are shown to arise from the states bound to each crossing of the waveguide, 
which is a photonic version of the {\it zero-point} effect in quantum mechanics and has not been recognized so far.  MPWNs thus make it possible to realize tight-binding bands without defects. 
The concept of the zero-point state turns out to be quite powerful in offering a clear understanding of low-lying eigenmodes in these systems. 
In terms of structures, three-dimensional MPWNs are similar to the 
inverted metallophotonic crystals of the woodpile type,\cite{lee2006three} 
but MPWNs cover a wider class of systems including two-dimensional structures as in Figs. 1(b) and 1(c). In those inverted PhCs, it has been recognized that air-band Bloch modes will localize to the air region. While this localization mechanism will not tell us whether the modes will localize to the crossings as in our results, or at "arms," the zero-point localization mechanism tells us that they will localize at the most vacant sites among the large dielectric constant region, namely, crossings due to the zero-point effect. Thus, the zero-point localization mechanism offers further insights into behaviors of eigenmodes than the conventional concept does.

 \begin{figure}[t]
 \includegraphics[width=8.5cm]{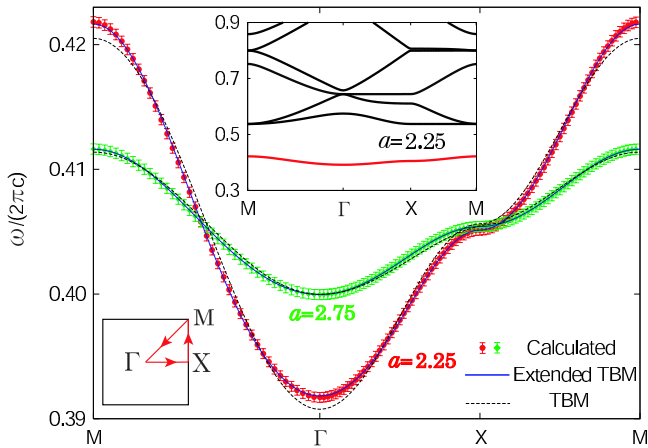}
 \caption{\label{a125_175}(Color online) Calculated photonic 
 band structure for the bottom band for a square-lattice 
 network [Fig. \ref{MPWN}(b)] for two values of $a$.  
 The error bars come from restricting the number of 
plane waves to 7225.  
 Curves represent tight-binding fits of 
the bands  with nearest-neighbor transfers (dashed lines) or 
with transfers extending to further neighbors (solid lines).  
Inset depicts low-lying bands for $a=2.25$. 
}
\end{figure}

Here, we shall  first show that the TBM is indeed a good approximation by 
taking a two-dimensional square-lattice MPWN as 
a typical example.  
To exemplify that we can use the tight-binding photonic bands 
to design the photonic band structures, we shall next consider 
metallophotonic waveguide on the kagome network. We shall 
show that the band structure for the kagome network contains 
{\it flat bands}, which are flat 
over the entire Brillouin zone with group velocities typically as 
small as 1/1000 times the velocity of light in vacuum. This 
amounts to a photonic 
realization of the ``flat-band model" in solid state physics as conceived by Lieb,\cite{PhysRevLett.62.1201} Mielke,\cite{mielke} and Tasaki.\cite{PhysRevLett.69.1608}   
Ultra slow light is among keen interests in PhCs,\cite{PhysRevLett.94.073903,vlasov2005acs} which has also been discussed in view of realizing a low-threshold laser. We then 
conclude that the tight-binding analogy should provide a systematic way to create photonic bands with desired band structures. 




{\it Band structure for a square lattice}. 
To demonstrate our basic idea for the tight-binding photonic band, 
we take the simplest example of a square-lattice network [Fig. \ref{MPWN}(b)] for 
the waveguide. Here, we focus on the TM-polarized mode (i.e., ${\bf E} \parallel z$) in two-dimensional networks. 
The validity of the TBM that we discuss here, however, can be extended 
to three-dimensional cases with arbitrary polarizations as mentioned later. While a band calculation for metallophotonic crystals is more complicated than those for PhCs with dielectrics, 
here we employ a simplified method as follows to extract 
the essence of the band structure. We assume that the metal that 
forms the wall of the network can be represented by 
a medium having a large and negative dielectric constant.  
Thus, we set the dielectric constant inside the waveguide to be unity, while 
the dielectric constant outside is taken to be $-10000$ 
in actual calculations. This should be adequate for describing a metal, at least in the microwave regime, 
where the metal can be effectively regarded as completely reflective. For 
shorter wavelengths, e.g., in the optical regime, the loss associated with the metal may not be ignored. A possible way out from the loss in this regime will be 
to employ line-defect waveguides since there is no loss associated with the material.
 
\begin{figure}[t]
 \includegraphics[width=7cm]{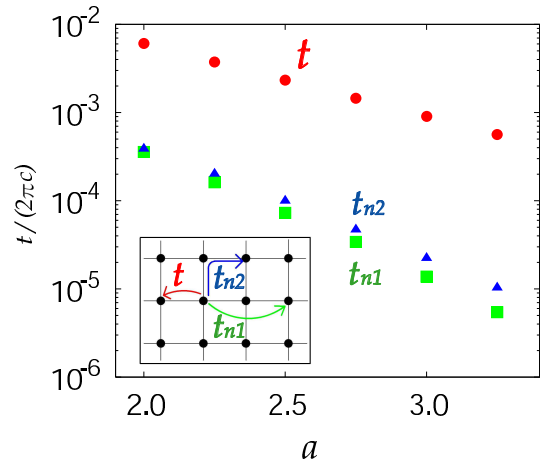}
 \caption{\label{box_exp}(Color online) The best-fit 
 ``hopping integrals" for the photonic tight-binding band 
plotted against the lattice constant $a$ for the nearest-neighbor transfer ($t$) and 
 further-neighbor ones ($t_{n1}, t_{n2}$) as depicted in the inset.  
 Errors in the fitting are smaller than the size of the symbols.  
}
\end{figure}

Band structures and eigenmodes have been obtained 
numerically with a plane-wave expansion method\cite{pendry1996cpb} for 
each value $a$ of 
the lattice constant of the network structure, while we set 
the width of each waveguide to be unity. We take a sufficient number of plane waves 
(typically 7225) to attain convergence.  

The band structure for the square-lattice network for $a=2.25$ is shown in the inset of Fig. \ref{a125_175}. We then fit the lowest band with the tight-binding band as shown in Fig. \ref{a125_175}  for two values of $a=2.25, 2.75$.   We can see that 
the fit is already good for the TBM with nearest-neighbor transfers, 
while the fit becomes even better when we include 
the further-neighbor transfers ($t_{n1}$, $t_{n2}$ in Fig.\ref{box_exp}).  
We quantify this by plotting the 
best-fit nearest-neighbor and further-neighbor transfers 
against the lattice constant $a$ in Fig. \ref{box_exp}. 
Each of the transfers is seen to decrease exponentially 
with $a$, where further-neighbor transfers decay more rapidly.

\begin{figure}[t]
 \includegraphics[width=7cm]{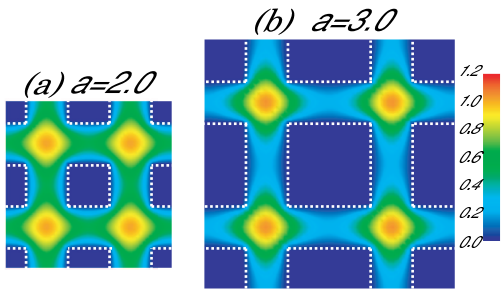}
 \caption{\label{Box_Eig}(Color online) Eigenmodes in the bottom band at $\Gamma$ point 
for (a) $a=2.0$ and (b) $a=3.0$. The boundaries of the waveguide are shown as dotted lines. Modes are normalized in each unit cell. 
}
\end{figure}

If we look at the eigenmodes, shown for the lowest band in Fig. \ref{Box_Eig}, they are in fact localized around each crossing, with a rapidly decaying tail toward the neighboring junctions.  We note that 
the existence of the bound amplitude localized around each crossing 
is analogous to the corresponding localized states in quantum 
mechanics, which exist due to a zero-point effect. Namely, for an {\it electron} system in crossed quantum wires, it 
has been shown\cite{PhysRevB.39.5476} that there exist bound states at the crossing.   
This comes from the fact that the wavefunction has a smaller zero-point energy at a crossing to form a bound state since the electron is less constrained there than in an arm. Schr\"{o}dinger's and Maxwell's equations for two-dimensional, TM-polarized PhCs are equivalent,\cite{PhysRevE.77.046602} so that the results in Schr\"{o}dinger's equation can be applied to the present system. When we translate the former 
to the present system, the lowest bound-state energy 
should be 0.406 (in units of $2\pi c$ with the waveguide width $=1$). The center of the tight-binding 
band (i.e., the on-site energy in the tight-binding fit) in Fig. \ref{a125_175} is $0.406\pm 0.001$, which excellently agrees with the present result.

Such an excellent agreement owes much to the formation of the bound states. By contrast, resonant states, which have intrinsic losses,  will couple to extended states to result in significant transfers to distant neighbors, deteriorating the tight-binding picture with near-neighbor transfers. The origins of the losses include the loss in the surrounding material, the out-of-plane energy leakage, and the radiation loss due to a finite size. While the first one may be avoided for metals in the microwave regime and for line-defect waveguides, the latter two losses can degrade the tight-binding fit.



We note that the existence itself of such localized states has been suggested 
for photonic crystal waveguides. Namely, it has been pointed out that there is a 
state in the photonic band gap with amplitudes localized around 
a T junction,\cite{PhysRevLett.77.3787} while a high 
transmission through a bend in a waveguide is observed for which a 
possibility for bound states is pointed out.\cite{albert2002pcm} However, the band structure due to a 
periodic array of such localized states in a periodic waveguide 
network has not been explored.

The tight-binding band considered here should emerge 
in general in MPWNs for arbitrary spatial dimensions and polarizations. Namely, the tight-binding bands emerge from the bound states via the zero-point analog states in MPWNs, which will 
continue to exist when we consider more complicated structures 
with arbitrarily polarized electromagnetic fields. Thus, we envisage here that 
MPWNs should generally accommodate tight-binding photonic bands.

There is a way to achieve tight-binding bands in existing studies employing coupled plasmon modes,\cite{PhysRevB.64.045117} but these bands should be susceptible to disorder in the surface of metals.  
The present system, on the other hand, should 
be robust against disorder. It has been known for cavities 
that a disorder in the surface of a cavity does not significantly decrease the $Q$ value.\cite{joannopoulos1999amw} The robustness comes from the fact that the field resides away from the surface, which also applies 
to the bound states considered here.

{\it Kagome lattice and a flat band}. 
To show that the present TBM can be a powerful guideline in 
designing photonic bands, we propose that the photonic 
band in a kagome network of waveguides [Fig. \ref{MPWN}(c)] should 
contain flat bands.  
A curious point is that the class of 
flat bands considered here is totally different from a dispersionless band 
in the narrow-band (i.e., zero-transfer) limit. The emergence of the flat band, originally considered in solid-state physics by Mielke and by Tasaki, is due to an {\it interference effect} inherently connected to the 
topology of the lattice, and thus the flat band exists 
no matter how large the transfers 
between the states at adjacent sites are.  
For electrons, there has been attempts to realize flat 
electronic bands in the kagome (a typical Mielke) network of quantum wires.\cite{shiraishi,Coulson}  So we pose a question here: can we realize the topologically 
originated flat photonic bands?

\begin{figure}[t]
 \includegraphics[width=7cm]{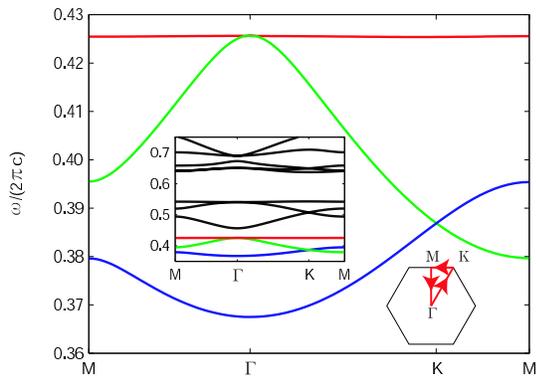}
 \caption{\label{kagome_band}(Color online) The lowest three photonic bands calculated numerically for the kagome network shown in Fig. \ref{MPWN}(c) for $a = 4.0$. Inset depicts a wider energy range.}
\end{figure}

The result in Fig. \ref{kagome_band} shows that the lowest three bands have the 
same forms as in the TBM, where 
the third photonic band from the bottom is indeed 
extremely flat, as expected.  The width of this band is in fact 
$3\times 10^{-4}$ (for $a=4$) in the present normalization, so that light becomes 
more than 1000 times as slow as light in vacuum. 
We note that the band becomes flat as an interference effect 
on the lattice where the nearest-neighbor transfer is primarily at 
work, so that resonant states with distant transfers would not 
realize this.

\begin{figure}[t]
 \includegraphics[width=7cm]{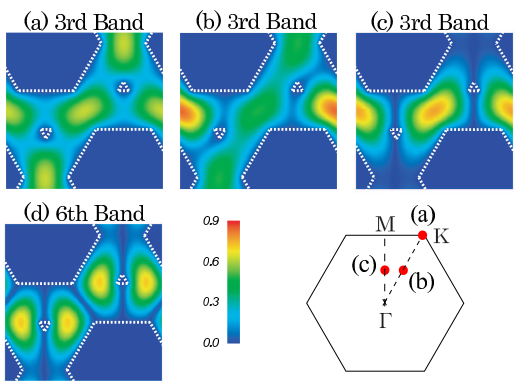}
 \caption{\label{eig_kagome}(Color online) Absolute value of the 
eigenmodes in the flat (third) band in a kagome network [Fig.1(c)] 
 with $a=4.0$ for several points (a)-(c) in the first Brillouin zone 
as shown in the inset.   
(d) is for the sixth band at point (c) in the Brillouin zone. Modes are normalized in each unit cell.}
 \end{figure}

It is intriguing to look at the eigenmodes in the flat band in Fig. \ref{eig_kagome}(a).   
While they are again localized around the crossing sites, 
the specialty of the flat band is appreciated if we look at 
the dependence of the eigenmode on the wave vector $\Vec{k}$.  
Namely, one hallmark of the flat band {\it \`{a} la} Mielke is that the $u_{\Vec{k}}$ 
part in a Bloch wave function ${\rm e}^{i\Vec{k}\cdot\Vec{r}}u_{\Vec{k}}$ 
strongly depends on $\Vec{k}$,\cite{PhysRevB.68.174419} while the eigenmodes in the zero-transfer limit obviously are not.   
We can confirm in 
Fig. \ref{eig_kagome} that the photonic eigenmode in the flat band indeed depends significantly on $\Vec{k}$ in the same manner as in 
the electron TBM on kagome.

Another notable feature is that there is more than one band which is flat. We can see that the sixth band 
is also flat as seen in the inset of Fig. \ref{kagome_band}, with the 
velocity less than 1/100 times the light in vacuum. If we compare the 
eigenmodes in the two flat bands in Figs. \ref{eig_kagome}(c) and \ref{eig_kagome}(d), 
we can see that the sixth band has an eigenmode with an 
extra node in the unit cell.  We can confirm that the two eigenmodes differ only in the nodal structure over the whole Brillouin zone. This implies that these two flat bands originate from different bound states with different symmetries.  The existence of more than one bound state is in fact 
also seen in an electron system 
at the jointing sites.\cite{PhysRevB.39.5476} 
Thus the photonic tight-binding bands can accommodate more than one set of bands corresponding to different symmetries in the bound-state modes.



In conclusion, we have considered MPWNs 
and shown that their band structures can be well captured as the 
tight-binding 
photonic bands. This phenomenon can be explained by bound states via the zero-point energy (in the language of Schr\"{o}dinger's equation) at the jointing sites. This offers a way to realize tight-binding bands in photonic crystals. To exemplify that the TBM can be exploited as a good guideline in designing photonic bands with desired band structures in MPWNs, we have shown that a kagome network accommodates flat bands with unusually small group velocities.  As a prospect there are a vast number of 
interesting lattice structures for MPWNs to be studied, among which are 
massless Dirac cones with the ``electron" (positive velocity) and ``hole" (negative velocity) branches in the honeycomb lattice.  
Other future problems include the detailed study of loss associated with metals for MPWNs in an optical regime. Quantitative comparisons with other systems and applicational details will also be future problems.

We wish to thank Kazuaki Sakoda and Takafumi Hatano for illuminating 
discussions.


\end{document}